\documentclass[Afour,sageh,times]{sagej}

\usepackage{moreverb,url}
\usepackage[dvipsnames]{xcolor}
\usepackage{listings}
\usepackage{graphicx}
\usepackage{float}
\usepackage{balance}

\definecolor{mGreen}{rgb}{0,0.6,0}
\definecolor{mGray}{rgb}{0.5,0.5,0.5}
\definecolor{mPurple}{rgb}{0.58,0,0.82}
\definecolor{backgroundColour}{rgb}{0.95,0.95,0.92}

\lstdefinestyle{PythonStyle}{
    language=Python,
    backgroundcolor=\color{backgroundColour},
    commentstyle=\color{mGreen},
    keywordstyle=\color{magenta},
    numberstyle=\tiny\color{mGray},
    stringstyle=\color{mPurple},
    basicstyle=\ttfamily\footnotesize,
    breakatwhitespace=false,
    breaklines=true,
    captionpos=b,
    keepspaces=true,
    numbers=left,
    numbersep=2pt,
    showspaces=false,
    showstringspaces=false,
    showtabs=false,
    tabsize=2
}
\newif\ifdraft
\draftfalse
\ifdraft
  \newcommand{\logan}[1]{}
  \newcommand{\ian}[1]{{\textcolor{blue}{ Ian: #1 }}}
\else
  \newcommand{\todo}[1]{}
  \newcommand{\logan}[1]{}
  \newcommand{\ian}[1]{}
\fi

\usepackage[hidelinks]{hyperref}

\newcommand\BibTeX{{\rmfamilyI  B\kern-.05em \textsc{i\kern-.025em b}\kern-.08em
T\kern-.1667em\lower.7ex\hbox{E}\kern-.125emX}}

\def\volumeyear{2016}

\begin{document}

\newfloat{code}{htbp}{lop}
\floatname{code}{Listing}

%\runninghead{Smith and Wittkopf}

\title{Employing Artificial Intelligence to Steer Exascale Workflows with Colmena}

\author{Logan Ward,\affilnum{1} J. Gregory Pauloski,\affilnum{2} Valerie Hayot-Sasson,\affilnum{2} Yadu Babuji,\affilnum{2} Alexander Brace,\affilnum{2} \\ Ryan Chard,\affilnum{1} Kyle Chard,\affilnum{2} Rajeev Thakur\affilnum{1} and Ian Foster\affilnum{1}}

\affiliation{\affilnum{1}Argonne National Laboratory, IL, USA\\
\affilnum{2}University of Chicago, IL, USA}

\corrauth{Logan Ward, Data Science and Learning Division, \\Argonne National Laboratory, Lemont, IL, USA}

\email{lward@anl.gov}

\begin{abstract}
Computational workflows are a common class of application on supercomputers, yet the loosely coupled and heterogeneous nature of workflows often fails to take full advantage of their capabilities. We created Colmena to leverage the massive parallelism of a supercomputer by using Artificial Intelligence (AI) to learn from and adapt a workflow as it executes. Colmena allows scientists to define how their application should respond to events (e.g., task completion) as a series of cooperative agents. In this paper, we describe the design of Colmena, the challenges we overcame while deploying applications on exascale systems, and the science workflows we have enhanced through interweaving AI. The scaling challenges we discuss include developing steering strategies that maximize node utilization, introducing data fabrics that reduce communication overhead of data-intensive tasks, and implementing workflow tasks that cache costly operations between invocations. These innovations coupled with a variety of application patterns accessible through our agent-based steering model have enabled science advances in chemistry, biophysics, and materials science using different types of AI. Our vision is that Colmena will spur creative solutions that harness AI across many domains of scientific computing. 
\end{abstract}


\maketitle

\section{Introduction}

Decades of steadily improving computer hardware have made computers often faster at answering questions than humans are at posing them. 
As such, artificial intelligence (AI) algorithms are playing an increasing role in science as both programmer and software. 
Supervised learning algorithms improving approximate models for costly simulations without human direction,
language models generating code that answers questions posed by humans as general questions, and many other marvels are commonplace. 
Under this context, the future of computational workloads may be filled with self-directed~software.

Workflows, applications that orchestrate execution of many diverse tasks, 
%composed of many similar tasks deployed concurrently, 
have been a major source of innovation in AI for high-performance computing (HPC).\citep{FerreiradaSilva2024workflowfrontiers}
The recurring nature of tasks provides the consistent, easily defined training sets that make AI integration simpler.
Key examples of AI in workflows include experimental design techniques that infer best inputs given the history of results \citep{jacobsen2018csp},
unsupervised learning techniques to draw inferences from output data on-the-fly \citep{lee2019deepdrivemd},
or generative techniques that invent what search spaces to explore \citep{GmezBombarelli2018vae}.
The growing range and increasing intelligence of AI models suggests these examples are an early example of a space filled with opportunity.

Optimal performance of the AI within an application relies on accounting for nuances in how it is used.
As an illustrative example, the AI tasks in an experimental design workflow only need to be performed when ``sufficient'' data are required,
and the notion of ``sufficient'' depends on many aspects of the application.
AI models that are fast compared to the tasks they advise could be run as each task is completed,
whereas relatively expensive AI models should be delayed and deployed on dedicated resources.
Expensive AI models may also benefit from a streaming policy where simulations are started based on intermediate results of the AI tasks, 
rather than waiting for all to complete.
Such ideas for harmonically composing simulation and AI tasks are still~growing.

We developed Colmena to allow scientists to create inventive solutions for integrating AI into workflow applications on supercomputing systems \citep{ward2021colmena}.
Colmena is a Python library that adds a layer to conventional workflow systems that simplifies expressing the dynamic ways AI can be used.
In this paper, we start by describing the previous work that inspired and enabled Colmena, then introduce its implementation before discussing a few HPC case studies.

\section{Related Work}

Our work sits at the intersection of AI and scientific workflows, providing a unique approach to combining them.

\subsection{AI Approaches for Science Workflows}

\logan{Convince readers that no single application type exists for AI+simulation}

There are numerous modalities for how AI can be used in simulation, and they can be understood by the relationship between AI and the simulation software it enhances \citep{fox2019learninghpc}.
The purpose of the AI could be to post-process the result of simulations and synthesize the outputs, as in the learning of collective variables in DeepDriveMD \citep{lee2019deepdrivemd, brace2022coupling}.
AI could also be used to create a fast surrogate that can be used to prejudge potential simulations from within a pre-defined search space \citep{stjohn2019mpnnpolymers}
or to generate potential simulations that resemble previous successes \citep{huerta2023ghp}.
Each different relationship implies a different order of operations in how the simulation and AI components are combined.

The strength of coupling between the tools and the relative degree of computational costs between AI and simulation impose further constraints on how applications are implemented.
Even within the archetype of ``AI used to create surrogates,'' distinct variations are possible. 
Applications may use the AI and simulation code concurrently, which requires low-latency inference from the ML models and therefore a need for dedicated resources for both computations \citep{caccin2015onthefly}.
In contrast, one that uses AI and simulation sequentially can use the same resources for both \citep{zamora2021proxima}.
Similar variations in degrees of coupling exist across other AI and simulation archetypes, yielding high diversity in the ways AI and simulation should be combined together.

\subsection{Workflow Engines}

\logan{Cover workflow engines, emphasizing how they're great at static workloads but not designed for dynamic workloads characteristic of AI}

Workflow engines and other related approaches to the task-parallel, often data-driven, execution of tasks on HPC systems has been a subject of research since at least the 1980s, when systems like Linda \citep{carriero1988applications} and Strand \citep{foster1989strand} were used for such purposes, while Condor \citep{thain2005distributed} and Condor-G \citep{frey2002condor} enabled dispatch of many tasks within one or multiple resource pools, respectively.
Parallel scripting approaches enable efficient execution of tasks coupled by file system operations \citep{zhao2005notation,wilde2009parallel}.

AI-enhanced applications provide challenges to the workflow engines that manage their computations on HPC.
The AI-based workflows that we consider here continuously update the tasks to be executed during the execution of the workflow. This is in contrast to the traditional static approach employed by many workflow engines that require the entire workflow graph to be defined a priori \citep{liu2015,herath2010streamflow}. 
Further, this dynamicity requires efficient workflow processing as the latency of tasks being created to running on a worker becomes as important as task throughput.
%AI continually determining new simulation tasks results in a stream of tasks being generated over time, resulting in the latency of task being created to running on a worker becoming as important as task throughput.
Tasks in AI-based workflows are particularly heterogeneous, with AI training tasks spanning many nodes while inference tasks require fractions of individual GPUs \citep{dhakal2023}. Such heterogeneity presents many challenges in efficiently scheduling tasks across available nodes \citep{phung2021het}.
It also creates new opportunities, for example, to deploy workflows 
%Heterogeneity motivates workflows 
that span multiple types of compute resources. 
%opening security and data movement issues.
In short, the addition of AI into scientific workflows will push the state-of-the-art on workflow engines.

\subsection{Integrating AI and Workflows}

\logan{Make sure readers know about libEnsemble, rocketsled, Ray, and SmartSim (at least). Explain how Colmena fits a different niche}

The variety of ways AI can be used in simulation has led to a bloom of approaches to implement them in software.
This emerging class of tools built to support workflows that include AI, which includes Colmena, share a common set of~features:

\begin{itemize}
    \item \textit{Templating for common design patterns}, such as how AI applications are broken down into generator, simulator, and allocator by libEnsemble \citep{hudson2022libensemble} or the problem definition templates of DeepHyper \citep{balaprakash2018deephyper}.
    \item \textit{Dedicated services or tasks for running AI tasks}, such as the persistent optimization server in CANDLE Supervisor \citep{wozniak2018supervisor} or the ``OptTask'' in RocketSled \citep{dunn2019rocketsled}.
    \item \textit{Integration with data fabrics}, such as via the use of SmartRedis to transmit data to AI tasks in SmartSim \citep{partree2022smartsim} or the two choices of data fabrics in eFlows4HPC.\citep{ejarque2022enabling}
    \item \textit{Connections to machine learning frameworks}, as illustrated by ties to PyTorch and Tensorflow within the hyperparameter tuning and training extensions for Ray \citep{mortiz2018ray}.
    \item \textit{Emphasis on dynamic allocation} to accommodate changes in tasks types as AI models improve, visible in the integration of dynamic schedulers in DeepDriveMD\citep{lee2019deepdrivemd}.
\end{itemize}

While similar along these dimensions, the tools vary significantly in how they define and deploy workflows.
For example, there is a divide between stateful processes with decentralized communication (e.g., SmartSim, Ray, Decaf \citep{yildiz2021decaf}) and others with a central controlling process (e.g., Supervisor, Colmena). 
Learning how best to grow these ranges of capabilities alongside the diversifying landscape of AI workflows remains an open question.

% \ian{We can cite SIMnet \citep{hennigh2021nvidia} and this paper \citep{lavin2021simulation} and this .
% How about \citet{zhong2021hybrid} and/or \citet{huerta2020convergence}.
% }

\section{Design}
\logan{Set the tone that Colmena is a way of composing workflows that achieves good performance because it is built on other libraries}

\begin{figure}
    \centering
    \includegraphics[width=\columnwidth]{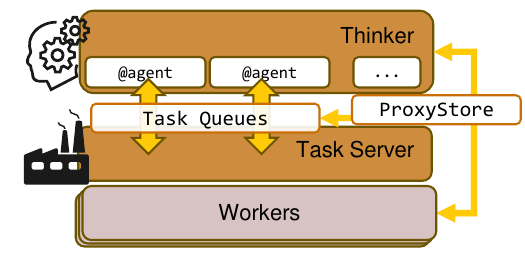}
    \caption{A Colmena application is composed of a Thinker and Task Server connected by a task queue.
    Thinkers define the policy for submitting computations using a series of agents that interact with each other and the Task Servers.
    Task Servers delegate computations to workers running on compute nodes. 
    Applications that manage large datasets or run at large scales can use ProxyStore to pass references to inputs and outputs via the workflow engine and object data via a side channel.
    }
    \label{fig:colmena-app-design}
\end{figure}

The purpose of Colmena is to write policies that schedule computation and data movement as Python functions.
We chose Python functions to allow policies of arbitrary complexity to be expressed in a well-known language.
Details for distributing computations are delegated to other libraries.
Colmena applications are composed of two components: a Thinker and Task Server (see application design,  \autoref{fig:colmena-app-design}).
We start by describing how policies are written as a Thinker and then detail how Colmena Task Servers use third-party tools to execute applications at scale.

\subsection{Programming Model}

\logan{Discuss how we define Thinkers and Task Servers}

A Thinker is a Python object whose methods define the policy of a computational campaign. 
Methods marked with special decorators run as threads after a user invokes the Thinker.
The threads, which we refer to as agents, submit computations to and receive results from a Task Server (described in detail in~\nameref{sec:design:execution}) over a shared queue.
Interactions between the agents and the results of tasks control how a computational campaign evolves.

\begin{code}
\begin{lstlisting}[language=Python,basicstyle=\ttfamily\small]
from colmena.thinker import BaseThinker
from numpy.random import random, sample
import numpy as np


class Thinker(BaseThinker):

    def __init__(
        self,
        queues,
        dimensionality: int = 8,
        num_samples: int = 256,
        walkers: int = 8,
    ):
        super().__init__(queues)
        self.n = num_samples
        self.d = dimensionality
        self.samples = []
        self.x = \
          sample((walkers, self.d)) * 2 - 1
        self.log_p = \
          np.zeros((walkers,)) + np.inf

    @agent(startup=True)
    def startup(self):
        for i, x in enumerate(self.x):
            self.queues.send_inputs(
                x, 
                method='compute_logp', 
                task_info={'w': i},
            )

    @result_processor()
    def step(self, result: Result):
        # Perform MC step
        w = result.task_info['w']
        new_lp = result.value
        old_lp = self.logp[w]
        accept = np.exp(new_lp - old_lp) \
            < random()
        if accept:
            self.logp[w] = new_logp
            self.x[w, :] = result.args[0]

        # Submit a new sample, if not done
        if not self.done.is_set():
            self.queues.send_inputs(
                self.x[w] + \
                  random((self.d,)) * 2 - 1,
                method='compute_logp',
                task_info={'w': w},
            )

        # Store, then stop if done
        self.samples.append(self.x[w])
        if len(self.samples) > self.n:
            self.done.set()

thinker = Thinker(queues)
thinker.run()
\end{lstlisting}
\caption{A Colmena Thinker that implements a parallel Metropolis-Hasting algorithm.
The Thinker stores the positions of each walker, the current probability for each position, and the output samples.
The \emph{startup} agent submits an initial set of computations then exits. 
The \emph{step} agent runs when a computation finishes, updates the state of the associated walker, then submits a computation for the next point.}
\label{lst:thinker}
\end{code}

Listing~\ref{lst:thinker} illustrates an example Thinker that implements a Markov Chain Monte Carlo sampling algorithm with two agents.
The \textit{startup} agent submits a population of tasks as soon as the run method is called and then exits, while the \textit{step} agent receives completed tasks and, for each, submits a new task so as to maintain a constant amount of work on the supercomputer.

Using Python functions to express the steering logic ensures room to design sophisticated strategies.
One could, for example, add a third agent that manages training a surrogate model and augment \textit{step} to use 
the surrogate model when reasonable, or introduce logic to restart sampling trajectories that become trapped in already-observed states, as in DeepDriveMD \citep{lee2019deepdrivemd,brace2022coupling}.
Colmena leaves avenues for optimization open.

\subsubsection{Agent Types} Colmena supports four types of agents that each fulfill common tasks in a workflow steering policy:
\begin{enumerate}
    \item \texttt{@agent} starts at the beginning of an application and is expected to run until the end of the application unless marked with a ``startup'' option.
    \item \texttt{@result\_processor} runs when a task of a certain type completes and is provided the result object (see \nameref{sec:tasks}) as an input.
    \item \texttt{@event\_processor} is invoked when an associated Python Event object is set.
    \item \texttt{@task\_submitter} executes when a certain number of resources are available. 
    Resources are tracked using a set of semaphores that can be accessed by all agents.
\end{enumerate}

\subsubsection{Threading}
We use Python's standard threading library to run agents in parallel and to coordinate between the agents.
For example, the resource tracking used by some Thinkers to balance the number of tasks of each type employs Python's built-in Semaphore objects.
Using standard libraries makes writing a Thinker as close to standard Python as possible.

The Global Interpreter Lock (GIL) of Python has yet to become a major limitation of Colmena applications.
The steering agents are intended to be lightweight, completing in just a few milliseconds for our case study applications \citep{ward2021colmena, ward2023hcw}.
Even if no agent may operate concurrently (i.e., if no parts of the operation release the GIL), a millisecond processing time per task places a general limit of thousands of actions per second---large enough for many applications.
We have considered adding the ability for some agents to run as separate Processes, which are free from GIL considerations but such a design would make it harder to coordinate with other agents.

\subsection{Defining Tasks}\label{sec:tasks}

The computations requested by Colmena Thinkers, \textit{Tasks}, are defined as Python functions.
As in other workflows, function definitions must be serializable (true for all functions defined in modules) and take inputs that can be serialized.
There is a great variety available within these bounds.
The tasks may be pure Python and run on a subset of a compute node, or make calls to external applications that span many compute nodes.
So long as they are defined via a Python interface, Colmena can run them.

The Thinker application requests a task using the name of the function and a series of positional or keyword arguments, as if calling the function locally.
The Thinker then prepares a Result object which captures the input information as well as any other information needed to define the task, such as resource requirements (e.g., a number of processors) or task metadata that would be useful in processing results later (e.g., an identifier connecting similar tasks).
Task Servers will populate the Result object with the results of the computation as well as communication overheads and execution times so that users can adjust subsequent tasks accordingly or, at least, analyze performance afterward.

\subsection{Task Queues}
\label{sec:design:queues}

The Task Queue communicates task requests and completed results between Thinker and Task Server.
A single application may use separate Task Queues for different classes of tasks so that groups of agents can operate independently. 
Applications can also use different Queue implementations.
Redis, for example, is well suited for large task rates or data sizes, but the complexity of running a Redis Server may not be justified in some cases, compared to Python's built-in Pipes.
All queues, regardless of type, use the same interface so that it is simple to exchange components to port an application between different computing systems or scales.

\subsection{Task Execution}
\label{sec:design:execution}

The Task Server stewards the execution of tasks requested by the Thinker.
The Task Server interface provides an abstraction over arbitrary workflow engines and is responsible for translating task requests from a Thinker, dispatching tasks to the workflow engine, and returning completed requests to the queue.
This abstraction ensures that Colmena applications are portable; the specific Task Server implementation can be exchanged allowing the same Thinker to run on different HPC systems.

Colmena provides Task Server implementations for Parsl \citep{babuji19parsl} and Globus Compute; other Task Server implementations can also be developed.
Parsl and Globus Compute can run arbitrary Python functions on arbitrary compute resources, from laptops to the largest supercomputers.
Parsl provides multiple types of executors suitable for different use cases and has shown to scale to workloads up to thousands of tasks per second.
Globus Compute is a cloud-managed, federated function-as-a-service platform.
The cloud-managed infrastructure enhances reliability and makes it simpler to deploy applications across multiple sites, in contrast to Parsl, which requires additional network configuration to use multiple compute resources concurrently.
Succinctly, Globus Compute trades task throughput, latency, and reliability for easy access to remote compute resources, although we note the performance tradeoffs are small for most workloads \citep{ward2023hcw}.

% \subsection{Separating Data from Control}
\subsection{Data Fabric}\label{design:proxystore}

\logan{Introduce ProxyStore and Globus}

Tasks in AI-centric applications often consume or produce copious amounts of data, which can lead to nontrivial communication overheads.
For example, all task data flows through the Task Server process in Colmena, so data-intensive tasks can result in heavy I/O burdens that slow down the process.
This challenge is not unique to Colmena as most workflow engines have a central coordinator, such as the Globus Compute cloud service, through which all task data must be transferred.
We provided tools in Colmena to circumvent these I/O bottlenecks.

Colmena integrates with ProxyStore \citep{pauloski2023proxystore,pauloski2024proxystore} to reduce communication overheads by moving data through specialized channels rather than through the Task Server and workflow engine.
ProxyStore replaces Python objects with \textit{proxies} that reference the location of the actual data and then resolve to the original object when used.
The proxy is small, making it suitable to transmit alongside the control messages of the workflow engine while the object data are propagated to workers using better-suited communication protocols (e.g., Redis, Globus, Remote Direct Memory Access).
In essence, ProxyStore translates task data from being passed-by-value to passed-by-reference, avoiding unnecessary copies of data across processes or expensive serialization. 
These proxies provide other benefits: proxies can be asynchronously resolved at the start of task execution to overlap compute and I/O, and I/O costs are not incurred for large objects when tasks exit early or fail unexpectedly.

Colmena can make use of ProxyStore in two different ways: by configuring
%ProxyStore can be used in two ways with Colmena: configure 
the \nameref{sec:design:queues} to automatically proxy large task objects and by proxying objects manually in the Thinker before task submission.
The two methods can be employed at the same time; in neither case do the Python functions comprising the Colmena tasks need to be modified.
% Colmena provides two ways for using ProxyStore, neither of which require the user to alter their task implementations.

Configuring the \nameref{sec:design:queues} to automatically proxy objects is the simplest way to obtain potential performance benefits.
In this approach, the user configures ProxyStore with parameters such as the communication protocols to be used, and passes that configuration to the queues.
Task objects (positional and keyword arguments) are then replaced automatically with proxies when a new task is created, and the result of a task is proxied automatically after execution.
%The queue requests ProxyStore to create a proxy of an object and store the serialized object in the secondary channel upon submission.
%The worker on a compute node resolves the proxy into the original object as the task is being run and then replaces the task's output with a proxy that is communicated back to the workflow engine.
However, this automated approach provides limited routes for optimizing transfer performance.
Thus it can also be beneficial to proxy objects manually, for example when an object is used by many tasks.
In this case, the user may create a proxy within the Thinker and pass that proxy as a task argument.

% The second route is to use ProxyStore directly in the agents of the Thinker process.
% Creating proxies explicitly allows many optimizations, such as reusing the same data across multiple tasks to take advantage of ProxyStore's caching
% or preemptively sending data to a compute resource before tasks are executed.
% As before, the actual data transfer happens without modifying the Python functions being executed by the Task Server.

Similar to how the Task Server abstraction makes it simple to redeploy a Colmena application on a different workflow engine or HPC system, ProxyStore decouples the configuration of communication protocols used to transmit task data from the application.
This reduces friction when migrating applications across systems with different networking or storage stacks.
No application code needs to be changed---only the ProxyStore configuration.
% ProxyStore removes the complexities of migrating applications across systems with different networking or storage stacks.

\subsection{Scaling on Supercomputers}
\logan{Capture what it takes to get close to Exascale}

Workflow engines such as Parsl permit most applications to scale to dozens of nodes without special effort.
The applications described in the section on \nameref{sec:case-studies} taught the Colmena development team many strategies for accessing scales in the hundreds or thousands of nodes:

\begin{enumerate}
    \item \textit{Passing data by reference} is critical for tasks that involve data larger than $O$(100)~kB or workflows that span more than one system \citep{ward2023hcw}. Passing large data via the workflow system leads to communicating tasks to compute nodes becoming a bottleneck. Object proxies are a powerful solution to such problems because task code need not be changed to resolve references, and caching accelerates tasks that reuse data, such as inference tasks that use the same model over many input batches.
    \item \textit{Avoiding unnecessary reinitialization} across functions can accelerate workflows. It is common for multiple tasks executed on the same worker to re-initialize the same expensive objects because workflow engines are designed to work with pure functions. In other words, the workers are not stateful actors. We circumvent this, for example, by keeping lookup tables or machine learning models used by tasks in RAM when not in use, rather than loading them from disk each time \citep{dharuman2023protein}.
    \item \textit{Acting on task completion rather than result reception} is possible in cases when a task finishing could inform the creation of a new task without the need to receive or process results yet. Employing ProxyStore to separate control messages and data flow means that result notifications can be received two orders of magnitude sooner, which can be exploited to hide the latency of data transfer \citep{ward2023hcw, harb2023lohc}. That is, we can act on a task finishing and defer processing the results until result data are available.
\end{enumerate}

\section{Case Study: Molecular Design}

\begin{figure}
    \centering
    \includegraphics[trim=3.5mm 3.5mm 3mm 2.5mm,clip,width=\columnwidth]{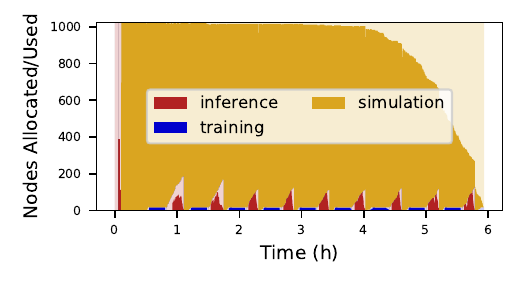}
    \caption{Allocation of HPC nodes between different tasks over time for a Colmena-based molecular design application. Nodes may either run quantum chemistry simulations (yellow), train a machine learning model (blue), or use the model to infer the properties of a molecule (red).  The application first runs inference on all nodes and then runs simulation tasks until sufficient data is available to begin re-training machine learning and re-running inference on a subset of nodes. Light shades indicate periods where either no computation was running or the running calculation did not complete before the end of the allocation. Figure from \cite{ward2021colmena}.}
    \label{fig:sc21}
\end{figure}

We used the design of molecules for redox-flow batteries as the prototype application for Colmena \citep{ward2021colmena}.
The application runs tasks that compute the performance of a molecule (i.e., solvation energy, redox potential),
train a model that predicts performance quickly,
or infer the performance of new molecules.
As in other examples of AI-driven design \citep{doan2020molbo,montoya2020camd,badra2022design,curtarolo2003csp,zhang2020dpgen}, our Colmena application achieves 
orders-of-magnitude improvements over unguided searches.
The advantage of Colmena is that we could access another 20\% increase in the number of high-performing molecules found by co-scheduling simulation and AI tasks (see Figure~\ref{fig:sc21}).
Achieving this increase in scientific performance while also maintaining effective use of the HPC resources required many innovations,
addressing challenges at different parts of the application.

\subsection{Reducing Communication Overheads}

\begin{figure}
    \centering
    \includegraphics[width=\columnwidth]{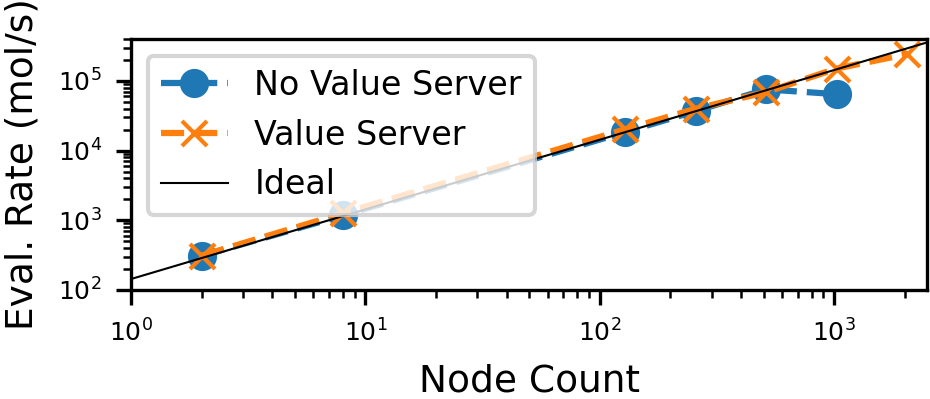}
    \caption{Weak scaling of inference rate as a function of node count for a molecular design application that uses message passing neural networks. Experiments were performed on the Theta Supercomputer at Argonne National Laboratory. Figure from \cite{ward2021colmena}.}
    \label{fig:value-server}
\end{figure}

The initial step in our molecular design application is to identify a set of target molecules by running inference over all molecules.
Inference tasks transmit large amounts of data (copies of the models, molecules, inference results) between many nodes,
which presents a clear challenge to scaling.
Large scales require faster transfer rates to keep all nodes populated with work and, as visible in 
Figure~\ref{fig:value-server}, data transfer became rate-limiting at only 512 nodes.

We identified the connection between the Task Server and worker nodes as the source of the problem.
Result data would remain on a compute node for as long as 10 seconds, which signals that the channels exchanging control messages between the workflow engine and workers have become saturated.
We alleviated the communication backlog by removing as much data from the control messages as possible and passing it instead through a ``Value Server'' (now available as ProxyStore \citep{pauloski2023proxystore}) that routes data directly between the Thinker and compute nodes.

We integrated ProxyStore into Colmena as part of the \nameref{sec:design:queues}.
The new queue intercepted inputs larger than a chosen size (10~MB in our case), stored their serialized representation in a Redis instance running on the same node as the Task Server, and replaced them with a reference.
A similar swap occurs on the compute node for result objects larger than a certain size.
Neither modifications to Colmena require changes to the source codes of the tasks and,
as shown in Figure~\ref{fig:value-server}, improved the scaling limit to above 2000~nodes.

\subsection{Using Specialized Hardware for AI Tasks}
Our AI-based optimization algorithm works best when the time between acquiring new data and updating recommendations for new simulations is minimized.
Updating recommendations requires retraining machine learning models and then rerunning inference, a task that requires about 45 minutes on the CPU-only nodes on the Theta supercomputer at Argonne National Laboratory  (see Figure~\ref{fig:sc21}).
We shortened this time by offloading machine learning computations to a GPU cluster, which required solving an amplified set of problems in data transfer \citep{ward2023hcw}.

We augmented the Colmena and Value Server framework used in \cite{ward2021colmena} to provide multi-resource compute and secure data transfer between resources via Globus.
An earlier version of Colmena required users to maintain SSH tunnels on at least three ports (two for the workflow engine, and one for the Value Store), which increases maintenance burden and may be disallowed on some systems.
Our next implementation used Globus Compute---previously known as FuncX \citep{chard20funcx}---to route task requests through a cloud service and Globus Transfer to move task data between systems.
Communicating tasks through Globus Compute required 100~ms and performing data transfer required at least 1~s: higher latencies than with our Parsl-based implementation but still small enough to mitigate.

We hid the data transfer latency by modifying the Thinker to use bulk transfers of task data in advance of tasks being executed.
Such optimized transfers are performed by manually creating proxies for task data rather than relying on the automated mechanisms introduced in \cite{ward2021colmena}.
Our molecular design app uses bulk transfer at several points: the molecules used in inference tasks once at the beginning of a run, the data used for training tasks each time model training starts, and the models used for inference as soon as model training completes.
Enacting the transfers manually reduces the number of times a transfer must be started (at a cost of 1~s each) and manually creating the proxy allows reusing them between tasks.
Reuse is visible in data transfer, accounting for less than 1\% of execution time for many inference tasks during a run.

\begin{figure}
    \centering
    \includegraphics[width=\columnwidth]{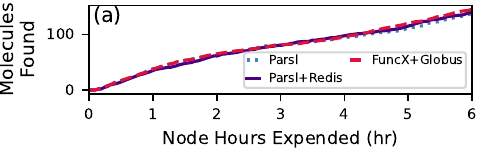}
    \includegraphics[width=\columnwidth]{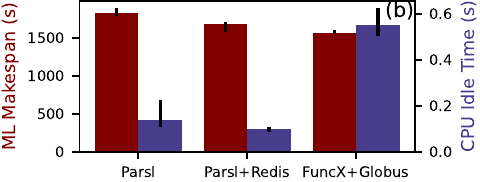}
    \caption{(a) Scientific output of our molecular design application over time and (b) key performance timings (time to complete machine learning tasks, average time between tasks for CPU workers) of a multi-site implementation of our molecular design application with different Colmena backends.
    Our implementations using the Parsl workflow engine and Parsl with Redis to transmit task data both required maintaining SSH tunnels between sites.
    The implementation with FuncX and Globus does not require direct network connections between sites, yet has similar scientific output and comparable performance timings.}
    \label{fig:hcw23-parslcomparison}
\end{figure}

The new data transfer coupled with the ability to hide latencies with tailored steering policies makes the convenience of cloud services available without performance penalties.
Figure~\ref{fig:hcw23-parslcomparison} shows that scientific output is unharmed by using a more resilient, Globus-Compute-based implementation, and the makespan of the machine learning tasks is even better. 
We attributed the increased performance in machine learning to the ahead-of-time transfer and noted that the increased latency in CPU idle time does not reduce overall utilization below 99\%.
In short, we made Colmena a suitable tool for workflows that can benefit from differing types of hardware for each task.

\subsection{Better Performance through Integrating More AI}
The large number of trailing tasks at the end of Figure~\ref{fig:sc21} signals an opportunity:
we can gain performance by breaking tasks into smaller parts.
Beyond the gains of just ensuring less information is lost at the end of an allocation,
introducing smaller tasks provides more opportunities for making smarter decisions with AI.
More decision points create both more tasks and 
sensitivity to latency in waiting for decisions,
leading to a stronger need for highly parallel computing systems.
We consequently are exploring finer granularity of AI and simulation tasks for the exascale version of our molecular design application.

We build multiple steps into our simulation workload by introducing multiple levels of fidelity.
Rather than compute the performance of a molecule at the target level of accuracy, as in \cite{ward2021colmena}, we now perform parts of the calculation incrementally (e.g., one property before another) and use multiple levels of accuracy for our simulation codes (i.e., smaller basis sets, cheaper DFT functionals). 
As demonstrated by \cite{woo2023pipelines} and \cite{reyes2022multistage}, 
these additional steps in accuracy reduce the cost of an optimization algorithm because
it is possible to stop evaluating low-performing candidates before incurring the full computational cost.

\subsection{Ongoing Challenges}
A few sources of performance degradation remain elusive. 

The under-utilization during the first minutes of Figure~\ref{fig:sc21} and during each subsequent batch of inference tasks is a result of the delay in loading Python libraries.
Speeding the library load rate can be accomplished by reducing the number of reads from the global filesystem \citep{kamatar2023python}.
We have yet to be able to make such solutions accessible to Colmena applications or the workflow engines that underlie them.

Individual molecule simulation tasks may involve launching an MPI executable multiple times, 
which can lead to significant overheads on large systems.
We have studied this problem using Colmena as a use case \citep{alsaadi2022rpmpi}
and intend to continue participating in the workflow community to adopt the latest advancements.

Ensuring each node running computations is used to its full extent is challenging because of the variety
of tasks in an AI workflow and potential variance in resource needs within tasks of a single type.
The molecular design application described here has served as a test case for 
exploring systems that identify automatically how to partition individual nodes for multiple tasks \citep{phung2021het} 
and for evaluating the effect of partitioning individual compute units within a node for tasks \citep{dhakal2023}.
We continue to investigate methods for node partitioning. 
%Node partitioning is a route we are continuing to build knowledge around.

\section{Other Successes}
\label{sec:case-studies}

We have implemented various applications using Colmena since our first prototypes in mid-2021 \citep{ward2023hcw,dharuman2023protein,harb2023lohc},
each highlighting different challenges and opportunities for braiding AI into workflows. 
We discuss two of these applications below.

\subsection{Protein Generation}

\logan{Emphasize where the start-up costs of tasks need special attention}

\begin{figure}
    \centering
    \includegraphics[width=\columnwidth,trim=1mm 0 0 0,clip]{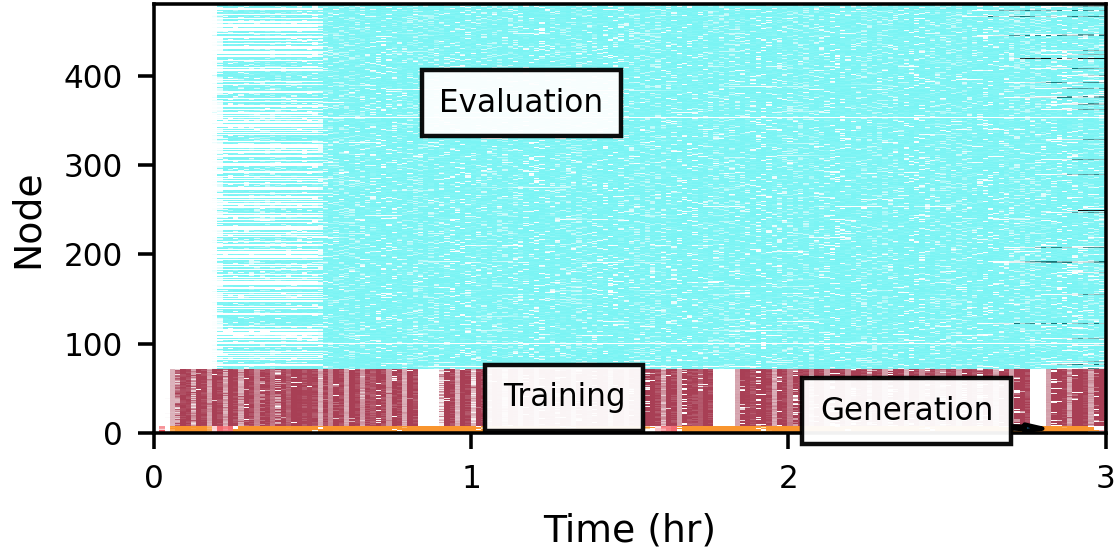}
    \caption{Utilization of each of 480 nodes (1920 GPUs) of ALCF's Polaris supercomputer over time for a Colmena application that simultaneously \textbf{trains} a reinforcement learning model that generates proteins, \textbf{generates} new proteins with that model, and \textbf{evaluates} the quality of the proteins. 
    Periods of higher utilization are represented as deeper shades and color indicates the type of task being run.}
    \label{fig:multirl-utilization}
\end{figure}

The large machine learning models central to the success of the protein design work of \citet{dharuman2023protein} were the primary source of scaling challenges.
The core of this application is a genome-scale language model (GenSLM) which produces genetic sequences that are then filtered to find the best-performing sequences \citep{zvyagin2023genslm}.
Then, another large language model \citep{lin2023esmfold} is used to fold the translated protein sequence, and a series of simulation steps are applied to the folded protein.
With Colmena, we could deploy all of these disparate task types together in a manner that balanced maximizing HPC utilization with scientific performance.

The central tradeoff of the protein design workflow is that generating or processing sequences without interruption maximizes system utilization, but more frequent reporting improves algorithm performance through better information flow.
As shown in \autoref{fig:multirl-utilization}, the largest periods of under-utilization in our application after the cold-start phase are the periods where the reinforcement learning training is being stopped and then restarted to update the version of the model used for generation.
Balancing this tradeoff required designing a Colmena application that lowered the cost of reporting through the following strategies:
\begin{itemize}
    \item \textit{Performing CPU-bound tasks asynchronously} from the GPU-intensive machine learning and simulation tasks. The Thinker application for Colmena was designed to process task results only after launching new tasks, ensuring that GPU tasks were dispatched with minimal delay.
    \item \textit{Caching large models in CPU memory} was necessary to allow tasks that both use large amounts of GPU models (e.g., protein folding, diversity scoring) to share the compute node. Keeping the model saved in memory increased the throughput of folding tasks by 30\% and prevented the nodes used for folding from being idled while downstream tasks were completed.
    \item \textit{Minimizing access to global filesystems} by having any task write intermediate data to local temporary storage (e.g., RAM disk) during computation, then passing completed results in-memory via ProxyStore rather than relying on the global filesystem to transmit results. Even though the data is serialized twice (once to local, once to ProxyStore), limiting the frequency of writes to the global filesystem is worthwhile.
\end{itemize}
A dynamic workflow system, such as Colmena, simplified expressing each of these strategies.

Our work with this application has allowed us to identify avenues for future improvements in Colmena.
Adding the ability for Colmena to report intermediate results would provide the largest improvement by
eschewing the need to restart tasks.
Integrating Colmena with in-situ workflow tools such as Decaf \citep{yildiz2021decaf} or 
with streaming systems such as Flink \citep{carbone2015apache} could allow for AI tasks to act as standalone services, while simulation tasks are served by a workflow system.
Expressing AI applications that blend workflows centered on atomic tasks and persistent services
is a research area ripe for exploration.

\subsection{Steering Molecular Dynamics}

\begin{figure}
\centering
\includegraphics[width=1\linewidth]{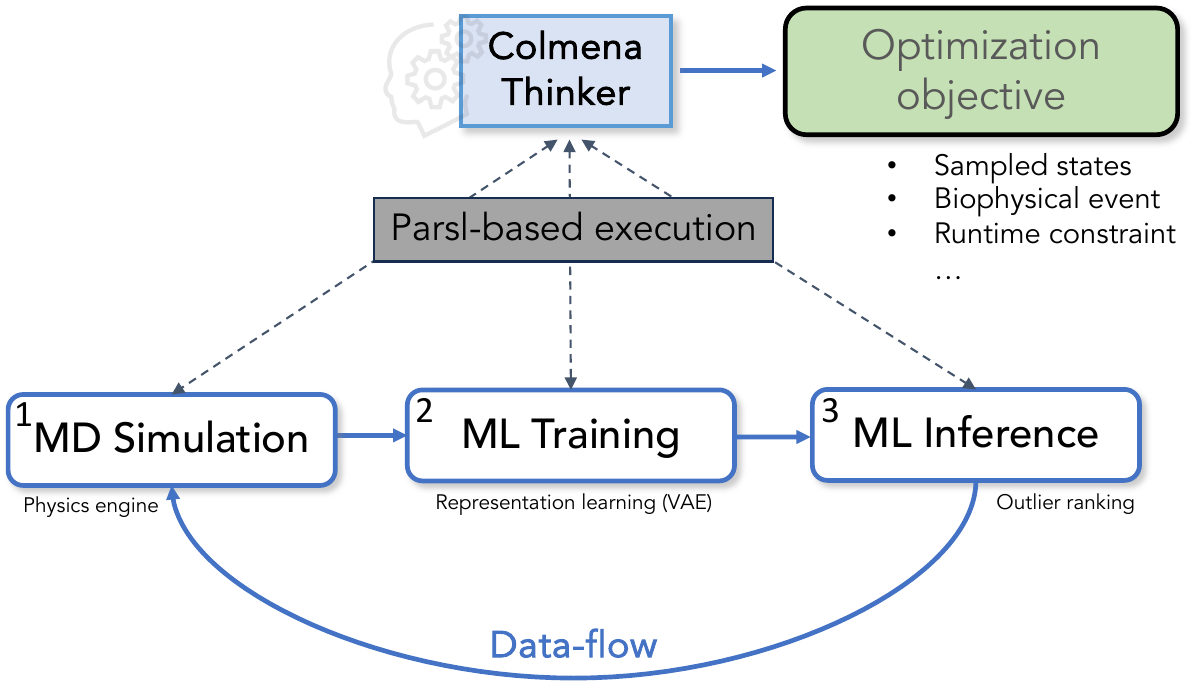}
\caption{DeepDriveMD accelerates the sampling of rare biophysical events within molecular dynamics (MD) simulations. A Colmena Thinker orchestrates an ensemble of MD simulation tasks while asynchronously training a machine learning (ML) model such as a variational autoencoder (VAE). ML inference and outlier detection are used to guide the simulation state sampling towards a customizable optimization objective.}
\label{fig:deepdrivemd}
\end{figure}

Molecular dynamics (MD) simulation of complex biomolecular systems is a prominent HPC application \citep{casalino2021ai, dommer2023covidisairborne, trifan2022intelligent, phillips2002namd} that acts as a computational microscope \citep{dror2012biomolecular} revealing biophysical details difficult to observe via experiment. Due to high free energy barriers, many important phenomena are difficult or impossible to sample using conventional MD, even with powerful supercomputers \citep{hospital2015molecular}. To approach this problem, \citet{lee2019deepdrivemd} and \citet{brace2022coupling} developed the DeepDriveMD framework, illustrated in  \autoref{fig:deepdrivemd}, for coupling ML/AI methods to MD simulations to track the simulated state space and guide simulations to sample more biophysically interesting events, constituting rare events.

The design pattern underlying DeepDriveMD involves steering an ensemble of many simulations using a trained ML model, such as a variational autoencoder \citep{bhowmik2018deep}, for inference. To keep the model up to date with the new data coming from the simulations, it must be periodically retrained. However, updating the model necessitates a speed versus accuracy tradeoff. Making immediate decisions on which simulations to stop or continue requires the use of a stale model (i.e., one that is not completely up-to-date). On the other hand, the delay induced by training may enable more accurate decisions that could better explore the simulation state space and ultimately lead to faster convergence for rare-event sampling.

DeepDriveMD, as a representative example of the Colmena steering paradigm, illustrates several broadly applicable strategies:
\begin{itemize}
    \item \textit{Asynchronous simulation and ML training} decreases the lead time for training ML models over a synchronized execution pattern, as the ML training is not blocked by simulation stragglers. In addition, hardware accelerators can be employed to further reduce the training time \citep{brace2021stream}.
    \item \textit{Streaming simulation data} for training and inference minimizes the file system I/O of the workflow, which is important for attaining high performance on leadership-scale facilities beyond 1,000 nodes.
    \item \textit{On-demand ML inference} allows the Thinker application to make decisions to stop or continue simulations without having to wait for the latest training task to finish, which leads to higher resource utilization and more MD sampling throughout a campaign.
\end{itemize}

While DeepDriveMD has shown up to $100$--$1000\times$ speed-ups for sampling protein folding pathways of certain biomolecular systems \citep{brace2022coupling}, we note that success is largely dependent on the proper alignment of the ML/AI method and the simulation data being generated. Hyperparameter tuning can play a large role in determining the success of sampling rare events, and domain-specific biophysical calculations are still needed to guide AI-driven sampling properly. Incorporating hyperparameter tuning frameworks such as DeepHyper \citep{balaprakash2018deephyper} may increase the robustness of AI-steered simulation workflows. Furthermore, uncertainty quantification through model ensembling \citep{egele2022autodeuq}, reinforcement learning adaptions \citep{shamsi2018reinforcement}, and incorporating statistically rigorous weighted ensemble approaches \citep{russo2022westpa} represent promising future directions for such workflows.

\section{Next Steps}

Our past and ongoing work building applications for Colmena has clarified a few routes for future development.

\subsection{Templates for Common Patterns}

We developed Colmena with the goal of giving application designers the freedom to write any steering policy, but we found many common elements that were repeated across applications.
One common example is an agent that submits the top task from a priority queue paired with an agent that updates the priority queue based on other completed computations.
Providing a library of templates will accelerate application development while ensuring access to well-tuned implementations of scheduling patterns.

\subsection{Integration with Model Repositories}

Machine learning approaches for even well-established problems are far from stagnant, 
which means the AI components of applications will be continually refreshed.
We plan to integrate Colmena with machine learning model repositories such as Hugging Face \citep{HuggingFace} and Garden \citep{Garden}
so that models can be treated as interchangeable components rather than hard-coded elements within an application.

\subsection{Intelligent Initialization}

The cost of reloading machine learning models has been a consistent challenge in developing Colmena applications, yet one we have only addressed with ad-hoc solutions.
A next step in Colmena development, in tandem with our workflow engine partners, will be to 
develop ``model registries'' or stateful actors that persist a shared state on workers between invocations of the same task.
Our initial prototype also provides mechanisms to define routes for clearing stateful objects as they become unneeded \citep{registery}.

\subsection{Streaming Intermediate Results}

Running some tasks as persistent services, rather than discrete tasks, will provide many
advantages including bypassing startup costs and dividing the costs of data transfer.
Introducing services will require breaking the assumption of workflow engines that functions are pure, so
a current and future aspect of research in Colmena is extending our programming 
model to support generator tasks that yield results continually without returning.
To achieve the required performance for deploying AI at larger scales, the generator tasks will need to be integrated within the data fabric (i.e., ProxyStore) as well.

\section{Proxy Application for Dynamic Workflows}

The core challenge of dynamic workflows, in our experience, is the ability to rapidly respond to the completion of tasks with new tasks.
The latency comes in three parts: a \textit{reaction} time between when a computation completes to when the Thinker is notified, a \textit{decision} time to produce the next time, and a \textit{dispatch} time for the new task to be delivered to a compute node.
We propose a single proxy application to determine the maximum scaling of a workflow system.

Our proxy application attempts to maintain a constant amount of tasks in the workflow.
The Thinker starts by preparing a list of computations then launching exactly as many as available workers, then launching a new task as soon as another is completed until the original queue of work is exhausted.
The tasks take an empty array as input, sleep for a duration drawn from a normal distribution, then return a random byte string.
The task rate of the workflow is varied by changing the worker numbers and the sleep duration distribution,
and the data communication costs are varied by altering the size of of the input and output data.

\begin{figure}
    \centering
    \includegraphics[width=\columnwidth,trim=3.5mm 5mm 3.5mm 3.3mm,clip]{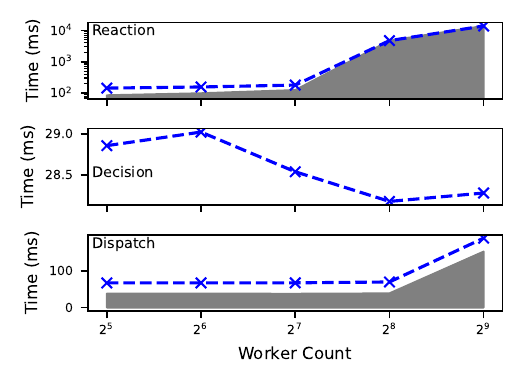}
    \caption{Latencies measured for a proxy application on up to 8 nodes of ALCF's Polaris supercomputer with task data sizes of 10~MB, a mean task length of 10~s, and a task length variance of 1~s.
    Each subplot shows a latency measure of particular relevance to dynamic workflows: 
    \underline{Reaction}: Task completion communicated from compute node to steering process; 
    \underline{Decision}: Steering process decides the next task;
    \underline{Dispatch}: New task delivered to idle compute node.
    In each case, the dashed line indicates the total latency and the shaded region the latency corresponding to latency without data transfer.
    }
    \label{fig:latency}
\end{figure}

Figure~\ref{fig:latency} shows the performance of our Proxy application on ALCF's Polaris supercomputer. 
We find that the major limit to applications for Colmena is latency of reacting to completed tasks, 
which becomes large after 256 tasks of approximately 10s each with data sizes of 10MB -- an approximate task rate of 25 tasks/second.
The fact that the latency increases with worker count helps identify a maximum sustainable task rate given
hardware and data sizes, which can be used to guide application design.
The proxy application narrows down that our performance could benefit from further tuning of data fabric employed ProxyStore,
or multiprocessing for processing completed tasks in Colmena.

As of Colmena v0.6.1, the ``task-limit'' applications are available as demonstration code in our GitHub repository.

\section{Conclusions}

We reviewed the inspiration, implementation, and impact of Colmena, a tool we designed as part of the Exascale Computing Project to explore routes for integrating AI into computational workflows on supercomputing systems.
Colmena allows scientists to describe workflow execution policies as Python functions
that schedule computations and data transfer on an HPC system.
We have employed Colmena successfully to implement applications that engage different types of machine learning (supervised, generative, unsupervised) in a variety of scientific domains---work that has both identified broadly applicable strategies for combining AI and simulation and suggested  priorities for further algorithmic and systems research.
We intend to continue simplifying the creation of new Colmena applications while exploring more routes for tailoring workflows to best use AI.

\begin{acks}
LW, GP, RC, RT, and IF acknowledge support by the ExaLearn Co-design Center \citep{alexander2021co} of the Exascale Computing Project (17-SC-20-SC) \citep{alexander2020exascale}, a collaborative effort of the U.S. Department of Energy Office of Science and the National Nuclear Security Administration, to develop Colmena and evaluate its performance on HPC.
YB and KC were supported to integrate Parsl with Colmena by NSF Grant 1550588 and the ExaWorks Project within the Exascale Computing Project.
GP, VHS, and KC were supported to develop ProxyStore by NSF Grant 2004894.
This research used resources of the Argonne Leadership Computing Facility (ALCF), a DOE Office of Science User Facility supported under Contract DE-AC02-06CH11357, including via the ALCF Data Science Program. It also used resources provided by the University of Chicago's Research Computing Center.
\end{acks}

\balance

\bibliographystyle{SageH}
\bibliography{colmena}

%\section{Support for \textsf{\journalclass}}
%We offer on-line support to participating authors. Please contact
%us via e-mail at \dots
%
%We would welcome any feedback, positive or otherwise, on your
%experiences of using \textsf{\journalclass}.

\end{document}